\documentclass[prl,aps,twocolumn,showpacs,showkeys]{revtex4}

\usepackage{graphicx}

\newcommand{\nxu}{n_{x,\uparrow}}
\newcommand{\nxd}{n_{x,\downarrow}}

\newcommand{\nyu}{n_{y,\uparrow}}
\newcommand{\nyd}{n_{y,\downarrow}}

\begin{document}

\preprint{DUKE-TH-01-213}

\title{Kosterlitz-Thouless Universality in a Fermionic System}
\author{Shailesh Chandrasekharan and James C. Osborn}
\affiliation{ 
Department of Physics, Box 90305, Duke University, \\
Durham, North Carolina 27708, USA }

\date{September 21, 2001}

\begin{abstract}
A new extension of the attractive Hubbard model is constructed to
study the critical behavior near a finite temperature superconducting
phase transition in two dimensions using the recently developed
meron-cluster algorithm. Unlike previous calculations in the
attractive Hubbard model which were limited to small lattices, the new
algorithm is used to study the critical behavior on lattices as large
as $128\times 128$. These precise results for the first time show that
a fermionic system can undergo a finite temperature phase transition
whose critical behavior is well described by the predictions of
Kosterlitz and Thouless almost three decades ago.  In particular it is
confirmed that the spatial winding number susceptibility obeys the
well known predictions of finite size scaling for $T<T_c$ and up to
logarithmic corrections the pair susceptibility scales as $L^{2-\eta}$
at large volumes with $0\leq\eta\leq 0.25$ for $0\leq T\leq T_c$.
\end{abstract} 


\maketitle

\section{INTRODUCTION}

A variety of critical phenomena in nature arise due to the development of
long range correlations or long range order at a finite temperature.
In the critical region
quantities are characterized by scaling relations that are universal
and depend only on the symmetries involved. The basis for this
universality is the renormalization group analysis \cite{Sta99}. A
number of universal properties have been confirmed with great
precision by numerically studying similar symmetry breaking patterns
in different microscopic models. However, almost all known examples
involve bosonic variables in the form of either classical \cite{Com00}
or quantum spins \cite{Bea98} at the microscopic level. On the other
hand, most interesting physical systems in condensed matter and high
energy physics fundamentally involve fermionic particles interacting
with each other through gauge fields. Unfortunately, calculations
confirming the predictions of universality in such systems are rather
limited and crude.  In particular precise calculations substantiating
the universal critical behavior in superconducting materials or
strongly interacting matter cannot be found.  The essential difficulty
is that the only known approach to study this subject is through
numerical simulations. In the case of Fermi systems one has to 
be clever and overcome the so called sign problem which makes designing 
algorithms difficult. Even in cases where the sign problem can be solved 
by integrating out fermions, the conventional algorithms
often suffer from critical slowing down. These difficulties have
restricted the calculations to small system sizes which are not
sufficient to accurately determine the scaling behavior near a phase
transition.

The lack of efficient numerical methods available for Fermi systems
has created some interesting controversies. It is well known that long 
range fermionic correlations can arise at zero temperature and can lead 
to novel universality classes \cite{Ros91}. On the other hand it is believed 
that the long range degrees of freedom near a finite temperature phase 
transition should be well described by bosonic degrees of freedom, the 
reason being that fermions acquire a mass proportional to the temperature 
$T$ and hence decouple from the critical behavior. However, a few years 
ago this conventional wisdom was questioned based on a large $N$ 
calculation and further substantiated by numerical simulations 
\cite{Koc95} at smaller $N$. It was suggested that the composite nature 
of the order parameter in fermionic systems may cause deviations from 
conventional universality. More recently a detailed study of the transition
revealed that the deviations could be related to the artifacts of the 
large $N$ limit \cite{Kog98}. Thus, although the controversy for the moment
appears settled the lesson one learns is that universality classes arising 
in fermionic theories can only be confirmed through precise non-perturbative 
calculations.

Recently a new method called the {\em meron-cluster algorithm}
has emerged as a very efficient alternative to solve certain classes of 
fermionic models \cite{Cha99.1,Cha99.2}. It is based on the well known
loop cluster algorithm for quantum spin models \cite{Eve93} and does not 
suffer from critical slowing down. Using the concept of a meron-cluster
it solves the sign problem \cite{Bie95}. This novel method has 
lead to the first successful determination of the critical behavior near 
a three dimensional Ising transition in a fermionic model, again confirming 
the predictions of universality \cite{Cha99.3,Cha00.1}. The calculations of 
\cite{Kog98} mentioned earlier, had reproduced the two dimensional Ising 
universality class from large (even) $N$ fermionic systems. 
This has now been extended to $N=1$ fermions using a meron-cluster 
method \cite{Cox00}. As far as we know, until now no one has been able 
to check the universality arguments with precision beyond the Ising
universality class starting from a microscopic fermionic model.
In this letter we present the first results from a 
meron-cluster algorithm which confirms that the critical behavior near
a finite temperature phase transition in a model consisting of 
fermions can be described by the Kosterlitz-Thouless (KT) universality 
class\cite{KT}.

The importance of the KT universality class is well known. A variety of
finite temperature phase transitions that arise in many two dimensional 
condensed matter systems involving phenomena like magnetism, 
superconductivity and superfluidity are expected to be described by this 
class. For example the superfluid transition in He$^4$ occurs due to a 
condensation of Helium atoms which in reality are tightly bound objects of 
fermionic constituents. Similarly the superconducting transition in high 
$T_c$ materials is expected to occur due to a condensation of electron pairs. 
Both these transitions are expected to follow the predictions of Kosterlitz
and Thouless and are related to a $U(1)$ particle number symmetry.
Although KT universality has been studied extensively using spin systems, 
it has been difficult to perform precise calculations confirming that a KT
universality can arise starting from a microscopic fermionic Hamiltonian. 
A simple model that has been studied in this context by conventional 
Monte Carlo methods is the attractive Hubbard model \cite{Sca89,Mor91}. 
Unfortunately the inefficiencies of the algorithm have limited the size 
of the systems that could be studied. This coupled with the fact that 
the model has an unusually low critical temperature and exponentially 
diverging correlation length have not yielded any quantitative information 
about the universality class of the transition \cite{Lac96}. Given 
the proliferation of transitions in nature that are conventionally expected 
to be described by the KT universality class, the fact that at a microscopic 
level most of the systems are always made up of fermionic degrees of 
freedom and the controversy discussed above regarding deviations from 
universality in Fermi systems, it is important to find precise calculations 
that either confirm our expectations or contradict them.

\section{The Model}

The fermionic system we study is a new extension of the attractive 
Hubbard Hamilton operator in two dimensions and is given by
\begin{eqnarray}
H &=& \sum_{<xy>}\;\Bigl[
\sum_{s={\uparrow,\downarrow}}
({c_{x,s}}^\dagger c_{y,s} + {c_{y,s}}^\dagger c_{x,s})
(1 - 3 n_{xy} + n_{xy}^2)
\nonumber \\
-&4&\left(\nxu -\frac{1}{2}\right)\left(\nxd -\frac{1}{2}\right)
\left(\nyu -\frac{1}{2}\right)\left(\nyd -\frac{1}{2}\right)
\nonumber \\
+&4& \left.
[\vec{S}_x\cdot\vec{S}_y +\vec{J}_x\cdot\vec{J}_y - J^3_x J^3_y]
\right]
\nonumber \\
-&4&\sum_x\;
\;\left(\nxu -\frac{1}{2}\right)\left(\nxd -\frac{1}{2}\right).
\label{hubbh}
\end{eqnarray}
Here $<xy>$ represents the nearest neighbor sites of a square lattice
of size $L\times L$ and $n_{xy}=\nxu + \nxd +\nyu + \nyd$. Although the 
Hamilton operator has many terms, it is easy to check that it is invariant
under $SU(2)$ spin transformations and conserves $U(1)$ fermion number. 
Further, there is an on-site attraction between electrons of opposite 
spins like in the attractive Hubbard model. As we will see, below a 
critical temperature transportation of fermion number through the bulk 
becomes easy, leading to superconductivity (or more 
appropriately superfluidity since the symmetry is not gauged in the model). 
In higher dimensions this is related to the spontaneous breaking of the
$U(1)$ fermion number symmetry. In two dimensions, since this is 
forbidden due to the Mermin-Wagner theorem, superconductivity occurs 
due to the KT phenomena.

The essential feature of the model that makes it useful is the fact
that its partition function can be written in terms of the statistical
mechanics of closed loops with positive weights \cite{Cha01}. To see
this we start with a discrete imaginary time approximation of the
partition function
\begin{equation}
Z = \mathrm{Tr}\left[\exp(-H/T)\right] \;\cong\; 
\mathrm{Tr}\left[ (\prod_{i=1}^4 [1 - \varepsilon\; H_i])^M\right]  
\label{pf}
\end{equation} 
where $\varepsilon = 1/(T M)$ and $H=H_1+H_2+H_3+H_4$ is a convenient
reorganization of the interactions present in the Hamiltonian. 
It is then possible to rewrite the path integral in terms of closed loops with
\begin{equation}
Z \;=\; \sum_{[b]}\; W[b] \;\;\overline{\mathrm{Sign}}[b]
\end{equation}
where $[b]$ is a configuration of bonds linking neighboring sites which
form closed loop clusters of sites. 
The magnitude of the Boltzmann weight is $W[b] >0$.
The fermion permutation sign is encoded in the topology of the loops.
When a configuration contains loops of certain topology, 
referred to as {\em merons}, then $\overline{\mathrm{Sign}}[b] = 0$. 
If there are no meron loops in a configuration then
$\overline{\mathrm{Sign}}[b] = 2^{N_{\cal C}}$ where 
$N_{{\cal C}}$ is the number of loop clusters in the configuration.
This novel representation of the partition function makes the model 
computationally tractable. More details on the meron-cluster formulation
can be found in \cite{Cha01}.

The approximate partition function given in eq. (\ref{pf}) becomes exact 
in the limit $M\rightarrow \infty$.
Although this continuous time limit can
be taken \cite{Bea96}, for simplicity the results presented here are 
obtained by fixing $M=20$. 
In the study of finite temperature critical behavior a
sufficiently large but finite $M$ is acceptable. This is due to the fact that 
critical dynamics arise from large spatial sizes and the discrete 
nature of the temporal direction only effects non-universal quantities. 
The two dimensional universal critical behavior near $T_c$ remains 
unaffected.

\section{Results}

The simplest observable relevant to superconductivity is the pair
susceptibility which we define as
\begin{equation}
\chi \; = \;
\frac{2 T}{Z V} \; \int_0^{1/T} \;dt\;
 {\rm Tr} \left[ \; {\rm e}^{-(1/T-t) H}\; p^+ \;
 {\rm e}^{-t H}\; p^- \;\right]
\end{equation}
with $p^+ = \sum_x c_{x,\uparrow}^\dagger c_{x,\downarrow}^\dagger$
the pair creation and $p^- = (p^+)^\dagger$ the pair annihilation operators.
The susceptibility contains information about the condensation of electron
pairs which are the on-site component of Cooper pairs in the BCS approach.
A formula for the susceptibility 
in terms of loop clusters is easy to construct using the results of 
\cite{Bro98}. In the present case, as discussed in \cite{Cha01}, the
susceptibility is proportional to the sum over the square of the size of 
certain clusters depending on the number of meron clusters in the
configuration. 
\begin{figure}[h]
\vskip0.3in
\begin{center}
\includegraphics[width=0.45\textwidth]{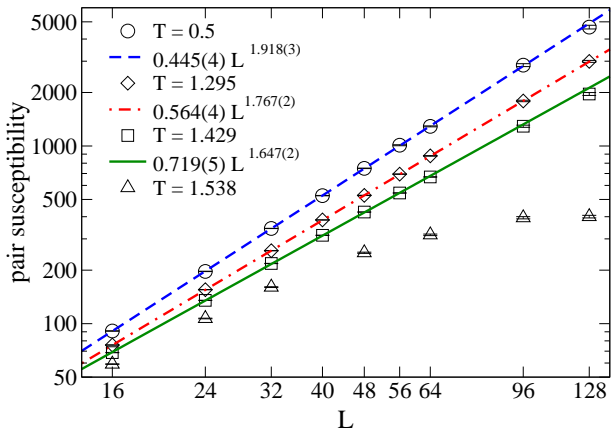}
\end{center}
\caption{\label{psus}
Pair susceptibility as a function of $L$.}
\end{figure}
In figure \ref{psus} we show results for the pair
susceptibility as a function of spatial size $L$. A state with 
quasi-long range correlations can be seen as a divergence in the 
susceptibility. Typically, for temperatures above the superconducting 
transition temperature $T_c$, the pair susceptibility should reach a 
constant for large enough volumes. This can be seen for $T=1.538$ and
$T=1.429$, although in the latter case lattices of size $L=128$ are
necessary to see the saturation suggesting that the correlation lengths
may be on the order of $100$ lattice units. This dramatic change in the 
correlation length between $T=1.538$ and $T=1.429$ is consistent with a 
Kosterlitz-Thouless prediction that it should diverge as 
$\mathrm{exp}(\mbox{Constant}/\sqrt{T-T_c})$ close to the critical 
temperature \cite{KT}. Interestingly, a fit to a power law over the 
range $L < 128$ is also poor at $T=1.429$.

Below $T_c$ the susceptibility should diverge as
\begin{equation}
\chi \;\propto \; L^{2-\eta(T)}
\label{fsspl}
\end{equation}
with the critical exponent $\eta$ starting at $1/4$ at $T_c$ and going down
to 0 as $T$ approaches zero. The exact formula
has a logarithmic correction to the power law which is small and cannot
be accurately determined with
reasonable size lattices so we ignore it here. This continuous change
in the power is also clearly visible. We find that $(2-\eta)$ is
$1.767(2)$ at $T=1.295$ and $1.918(3)$ at $T=0.5$. The fits to a power law
are extremely good for both these temperatures over the entire range of $L$.

\begin{figure}[h]
\vskip0.3in
\begin{center}
\includegraphics[width=0.45\textwidth]{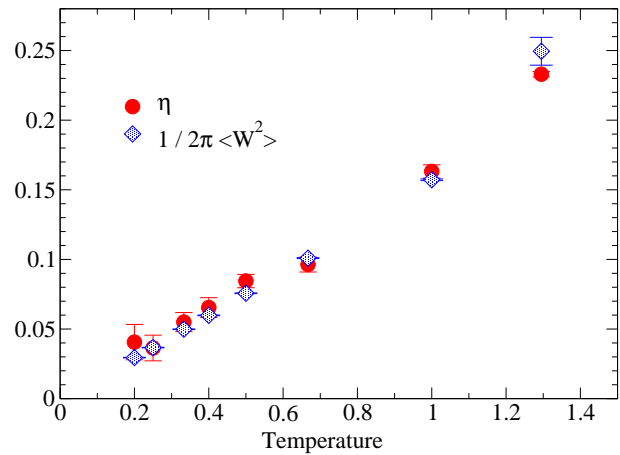}
\end{center}
\caption{\label{wnsus}
Comparison between $\eta(T)$ and $1/2\pi \langle W^2\rangle$ as
a function of $T$.}
\end{figure}

Although the pair susceptibility alone is sufficient to test for
the Kosterlitz-Thouless predictions, there could always be a 
lingering doubt whether the power law fits would fail
when one goes to much larger lattices. For example if we only
had data for $L\leq 32$, a power law fit could even work well
for $T=1.429$ when the error bars are sufficiently large. Larger
lattices were crucial to determine that the power law was not
a good fit in that case. In order to alleviate such worries we 
looked at the winding number susceptibility which we define as
\begin{equation}
\langle W^2\rangle = \left< \; (W_{\mathrm{x}}/2)^2 + (W_{\mathrm{y}}/2)^2 \;
 \right> /2
\end{equation}
where $W_{\mathrm{x}}$ ($W_{\mathrm{y}}$) is the total number of fermions
winding around the boundary in the x (y) direction.
This quantity is very useful in
measuring $T_c$ and has been used in bosonic systems \cite{Pol87}.
Although it is a difficult quantity to measure with conventional
algorithms, it is relatively easy in the meron-cluster approach showing the
power of the new method. Further, we know its finite size scaling form 
to be
\begin{equation}
\pi\;\langle W^2 \rangle 
~=~2+\sqrt{\Delta(T)}~\coth(~\sqrt{\Delta(T)}~\log(L/L_0(T))~),
\label{fsshm}
\end{equation}
with $\Delta(T_c)=0$ \cite{Har98}. It can be shown that for  $T < T_c$ in the 
$L \rightarrow \infty$ limit $2 \pi \eta(T)\langle W^2\rangle = 1$
\cite{Nel77,Pol87}. The fact that $\langle W^2 \rangle$ jumps to the
universal number $2/\pi$ from $0$ is another well known feature
of the KT phenomena and can be used to determine $T_c$.
Since $\eta(T)$ can be determined from the scaling 
of the pair susceptibility one can combine it with the measurement of 
$\langle W^2\rangle$ below $T_c$ to check for consistency in the KT 
universality class.
Figure \ref{wnsus} compares $1/(2\pi \langle W^2\rangle)$ obtained by 
extrapolating the values of the winding number susceptibility to the 
infinite volume limit using the formula (\ref{fsshm}) and $\eta(T)$ 
obtained from the finite size scaling of the pairing susceptibility,
as a function of temperature. Clearly the results are in 
excellent agreement with the predictions for a KT phase transition.

\section{Directions for the Future}

This work can be extended in several directions. One interesting
problem in condensed matter physics is to understand how disorder
effects superconductivity. It is predicted that at zero temperature
as the disorder is increased the system undergoes a quantum phase 
transition to an insulating phase\cite{Gol98}. If this is indeed the 
case, it would be 
interesting to understand the underlying critical behavior starting
from a fermionic theory. Previous studies have used the attractive 
Hubbard model as the starting point \cite{Sca99}. We suggest
that the model studied here is perhaps a better alternative since
the meron-cluster algorithms could turn out to be more efficient in such
a study.
Many other fermionic models with continuous symmetries
and applications in condensed matter, nuclear and high energy physics
can be studied with meron-cluster algorithms.
A systematic approach has been outlined in \cite{Cha01}. In particular 
it has also been shown that in certain regions of the parameter space
some repulsive Hubbard-type models with a non-zero chemical potential 
can be solved with meron-cluster techniques. Whether there is
d-wave superconductivity in these models is an open question. A 
common feature of these new models is that they appear to be more 
complicated. However since we have algorithms for them which give a 
computational advantage, they may still produce the clearest results. 
It is a common practice to work with models that are the easiest to solve
(numerically in this case) to help understand complicated physical systems.
Once the basic phenomena is understood one can then attempt more difficult
models to explore the robustness
and variations in the details of the underlying physics.

\section*{Acknowledgments}

We thank Uwe-Jens Wiese and Harold Baranger for many fruitful discussions. 
We also thank Xincheng Xie and Richard Scalettar for clarifications about 
the physics of the Hubbard model. This work is supported in part by funds 
provided by the U.S. Department of Energy grant DE-FG02-96ER40945 
and the National Science Foundation grant DMR-0103003.
The computations were performed on {\bf Brahma}, a Pentium based 
Beowulf cluster constructed using computers donated generously by 
the Intel Corporation and located in the physics department at Duke 
University.

\end{document}